# A Design Theory for Transparency of Information Privacy Practices

**Authors**: Tobias Dehling, Ali Sunyaev; Institute of Applied Informatics and Formal Description Methods, Department of Economics and Management, Karlsruhe Institute of Technology, Karlsruhe, Germany and KASTEL Security Research Labs, Karlsruhe, Germany; {dehling, sunyaev}@kit.edu

**Abstract**: The rising diffusion of information systems (IS) throughout society poses an increasingly serious threat to privacy as a social value. One approach to alleviating this threat is to establish transparency of information privacy practices (TIPP) so that consumers can better understand how their information is processed. However, the design of transparency artifacts (eg, privacy notices) has clearly not followed this approach, given the ever-increasing volume of information processing. Hence, consumers face a situation where they cannot see the 'forest for the trees' when aiming to ascertain whether information processing meets their privacy expectations. A key problem is that overly comprehensive information presentation results in information overload and is thus counterproductive for establishing TIPP. We depart from the extant design logic of transparency artifacts and develop a theoretical foundation (TIPP theory) for transparency artifact designs useful for establishing TIPP from the perspective of privacy as a social value. We present TIPP theory in two parts to capture the sociotechnical interplay. The first part translates abstract knowledge on the IS artifact and privacy into a description of social subsystems of transparency artifacts, and the second part conveys prescriptive design knowledge in form of a corresponding IS design theory. TIPP theory establishes a bridge from the complexity of the privacy concept to a metadesign for transparency artifacts that is useful to establish TIPP in any IS. In essence, transparency artifacts must accomplish more than offering comprehensive information; they must also be adaptive to the current information needs of consumers.

# INTRODUCTION

Information systems (IS) increasingly permeate society (Yoo 2010) and evolved from tools that *reflect* reality to tools that *shape* reality (Baskerville et al. 2020, Demetis and Lee 2018). The rising volume of information processing throughout society, as well as the potential of IS to reshape reality, poses an increasingly serious threat to privacy as a social value (DeCew 1997) and a pillar of functioning democracies (Schwartz 1999). Due to the ubiquity and complexity of information processing, consumers cannot see the 'forest for the trees[1]' when using IS and are no longer able to ascertain whether information processing meets their privacy expectations. In this manuscript, we present an IS design theory (ISDT) that establishes a theoretical foundation for IS that are less prone to undermine privacy as a social value by establishing transparency of information privacy practices (TIPP). In other words, TIPP theory explains and prescribes what transparency artifacts should be built to reveal the information consumers need to interact with IS in line with their privacy expectations.

Designing transparency artifacts (ie, artifacts for establishing TIPP) entails navigation of a trade-off between under- and overprovisioning of information: If transparency artifacts do not offer the information of interest to consumers, they are not useful for establishing TIPP because they cannot satisfy consumers' privacy information needs. On the other hand, transparency artifacts will also not be useful for establishing TIPP, if they present more information than consumers can cognitively handle. This will lead to a situation of privacy fatigue where consumers are overtasked with privacy management and may stop taking their privacy expectations into account when interacting with IS (Agozie and Kaya 2021). TIPP theory serves as theoretical foundation how this trade-off can be navigated with artifact designs that are adaptive to consumers' evolving privacy expectations and corresponding information needs, while simultaneously avoiding underprovisioning of information.

---

[1] We use this metaphor in the sense that persons are in a situation where they cannot see in which situation they actually are. The forest they would like to see represents a state where they can determine whether the privacy practices in an IS are appropriate with respect to their privacy expectations, but they are confronted with so many information on privacy practices (representing the trees) that this is just impossible.

Commonly instantiated transparency artifacts are privacy notices (Slepchuk and Milne 2020). Privacy notices are notices posted on websites that offer background information on the provider and introduce the privacy practices in the IS (Pollach 2006). Given the current state of ubiquitous information processing, privacy notices have largely outlived their usefulness for establishing TIPP. Consumers have neither the knowledge, time, nor the systems available to painstakingly manage privacy (Obar 2015). Already in 2008, it was estimated that each US citizen would spend between 181 and 304 hours per year to read (or between 81 and 293 hours per year to skim) privacy notices for each new website they visit (McDonald and Cranor 2008). Some progress has been made to improve privacy notices—for instance, by improving the information offered (eg, Reidenberg et al. 2016, Sánchez et al. 2021) or the user interfaces for communicating the information (eg, Karegar et al. 2020, Schaub et al. 2017). Still, it is unlikely that privacy notices can be improved far enough to become useful transparency artifacts. Expecting consumers to read privacy notices is just an unreasonable demand given the ubiquity of information processing in this day and age.

The situation is akin to the underwhelming outcomes when it was tried to introduce factory work processes and increase mechanization in mines in the United Kingdom in the 1950s, which led to the emergence of sociotechnical thinking (Trist 1981). In contrast to expectations, mechanization of mining led to decreased productivity and lower worker satisfaction due to poor alignment between factory work processes and work conditions faced in mine environments (eg, frequent unexpected events or hard-to-oversee and evolving mine shafts; Trist 1981). Bostrom and Heinen (1977) also called for more attention to emergent interactions and better alignment between technical characteristics of IS and demands of the social (organizational) environments, in which management-focused IS are supposed to operate, to improve artifact design. A more recent literature review also came to the conclusion that the sociotechnical perspective yields value by generating insights how to better align achievement of instrumental (eg, data protection or information security) and humanistic (eg, privacy or safety) objectives of IS (Sarker et al. 2019). To do so with respect to transparency artifact designs useful for establishing TIPP, we approach design of transparency artifacts from a sociotechnical perspective. Rather than restricting our theory development to the design logic of privacy notices, we account for consumers' privacy expectations in the design of transparency artifacts by examining the issue top-down from the perspective of privacy as a social value.

In particular, we do not rely on the assumption that consumers' privacy behaviors result from a privacy calculus, which is a key assumption underlying the design of privacy notices (Hoofnagle and Urban 2014). Performing a privacy calculus implies that consumers maximize benefits and minimize risks when engaging in privacy behaviors (Dinev and Hart 2006). The assumption that consumers perform a privacy calculus for each of their myriad interactions with IS has become a rather utopian idea because the volume of information processing is constantly increasing (Yoo 2010) whereas consumers' cognitive capacities are not (Miller 1956). Moreover, uncertainties about consequences and preferences regarding privacy behaviors, as well as context-dependent changes in privacy expectations and susceptibility to manipulations by third parties in today's IS, largely prevent consumers' from performing such privacy calculi (Acquisti et al. 2015, Dinev et al. 2015).

TIPP theory also allows for privacy behaviors that are context-dependent (Nissenbaum 2010) and largely informed by environmental characteristics and personal privacy norms (Bélanger and James 2020). Under consideration of context-dependent privacy behaviors, it is unnecessary and even counterproductive to reveal all the information required to perform a privacy calculus. TIPP theory suggests that establishing TIPP is about revealing the information consumers need to perform privacy behaviors. Our goal is to encourage more useful designs of transparency artifacts by approaching their design top-down from a social privacy perspective. A stronger link between privacy as a social value and design knowledge for transparency artifacts will be helpful to design artifacts that can account for consumers' context-dependent privacy expectations.

We present TIPP theory in two parts to account for the sociotechnical interplay. The first part conveys design-relevant explanatory/predictive theory (DREPT; Kuechler and Vaishnavi 2012), which translates abstract knowledge regarding the IS artifact and privacy into a description of social subsystems of transparency artifacts. The second part conveys prescriptive design knowledge in form of an ISDT focused on the design product (Walls et al. 1992, 2004) to outline a solution space for technical subsystems of transparency artifacts, which is grounded in kernel theory. The main contribution of TIPP theory lies in establishing a bridge from the complexity of the privacy concept to a solution space for transparency artifacts that are useful for establishing TIPP. TIPP theory yields insights for any IS[2] because it informs designers regarding what should be built to avoid designs of transparency artifacts where consumers cannot see the 'forest for the trees'.

In the next section, we ground the motivation for our research in extant literature on the IS artifact and privacy. Afterward, we delineate how we developed TIPP theory based on Weick's framework of disciplined imagination (1989).[3] Our presentation of TIPP theory starts with the DREPT-part of the theory (Kuechler and Vaishnavi 2012) and concludes with the ISDT-part (Walls et al. 1992, 2004). Finally, we conclude with a discussion of the limitations and implications of TIPP theory.

## INFORMATION SYSTEMS AND PRIVACY

To set the stage, we first present background information on the interplay of social and technical subsystems of IS artifacts. Next, we discuss extant research on transparency and privacy to position TIPP theory within existing research.

### Subsystems of IS artifacts

IS artifacts are sociotechnical systems that consist of technical and social subsystems, which should be well aligned to account for their emergent interactions and attain the instrumental as well humanistic objectives of IS (Sarker et al. 2019). Technical subsystems consist of hardware, software, and techniques predominantly designed and configured to reach instrumental goals (Chatterjee et al. 2020). Social subsystems encompass the "relationships or interactions between or among individuals through which an individual attempts to solve one of his or her problems, achieve one of his or her goals or serve one of his or her purposes"[4] (Lee et al. 2015, p. 9). IS artifacts emerge from the interactions of their subsystems (Lee et al. 2015), and IS artifacts thrive if technical subsystems are well-aligned with social subsystems (De Leoz and Petter 2018). Misalignments between technical and social subsystems can be corrected by adapting the technical or social subsystems to better fit each other (De Leoz and Petter 2018).

---

[2] Since our work is based on the IS artifact conceptualization by Chatterjee et al., *any IS* refers to any IS that can be "represented as a superordinate system composed of social and technical subsystems" (2020, p. 556).

[3] The key rationale for building upon Weick's framework of disciplined imagination (1989) to develop TIPP theory is the methodological freedom that it affords for theory development. Approaching design of transparency artifacts top-down from the perspective of privacy as a social value demanded attention to the abstract, intangible, and wicked concept of privacy as a social value as well as more tangible and tame technological concepts; methodological freedom is quite helpful to deal with diverse concepts. Methodologically more constrained approaches would most likely miss out on pieces of the puzzle. This manuscript is focused on the presentation of the final version of TIPP theory. Please refer to the appendix for examples of thought trials and their implications for the development of TIPP theory.

To improve the design of transparency artifacts, better alignment between the technical and social subsystems primarily implies adaptation of technical subsystems. Social subsystems are more relevant to guide and constrain the development of transparency artifacts, because it is hard to predict whether changes to social subsystems are more likely to strengthen or subvert privacy as a social value. Redesigning social subsystems (eg, by passing a law requiring consumers to read privacy notices, even if they are not interested in doing so) can result in inappropriate interference with consumer behaviors, which would constitute a privacy violation in itself (Marmor 2015, Solove 2006).

This corresponds with Simon's idea (1996) that at the heart of design is a thin interface between the inner and outer environment of an IS. The inner environment (technical subsystem) of an IS "is the hardware and operating ecology, commonly referred to as the computer system infrastructure" (Niederman and March 2012, p. 1:2). The outer environment (social subsystem) of an IS "is the people, organizations, and societies served by the information system" (Niederman and March 2012, p. 1:2). "If the inner system is properly designed, it will be adapted to the outer environment, so that its behavior will be determined in large part by the behavior of the latter" (Simon 1996, p. 11–12).

From a privacy perspective, the challenge of establishing alignment between the inner and outer environments of IS artifacts and to account for their emergent interactions is that privacy is a social value (DeCew 1997). Due to the plurality of goals that have to be considered when dealing with social issues, it is usually not possible to abstract the outer environment into a clear goal formulation for the inner environment (Rittel and Webber 1973). As a technical notion, privacy can be thought of as an incomplete requirements specification with requirement weights that differ between different individuals and change over time, depending on the context of the individual, because of the pluralistic, evolving, and contextual nature of consumers' privacy expectations (Mulligan et al. 2016, Nissim and Wood 2018). Social issues constitute *wicked problems*; thus, such problems demand attention to constraints imposed by social subsystems and cannot be resolved through mere optimization of technical subsystems (Rittel and Webber 1973) because unknown requirement weights or entirely unknown requirements cannot be purposefully and satisfyingly accounted for in technocentric approaches. This is in contrast to *tame problems* (eg, development of an encryption algorithm), where goals (eg, protect confidentiality, limit computational overhead, achieve quantum safety) can be clearly specified and reached by optimizing technical subsystems. Addressing wicked problems, which include privacy problems, encompasses consideration of the emerging interactions between social and technical subsystems of IS. In the next section, we discuss social subsystems of IS artifacts in more detail from a privacy perspective.

---

[4] We would like to note that this definition of social subsystems accounts not only for the needs of the individual but also for the needs of other social actors that emerge from interactions and relationships between individuals (eg, organizational actors like IS providers).

**Privacy and IS artifacts**

Privacy is an essentially contested concept that is addressed in research from diverse perspectives (Mulligan et al. 2016). On an abstract level, these perspectives can be grouped into legal, technical, psychological, and social perspectives on privacy. The legal perspective focuses on data protection legislation, for example, the California Consumer Privacy Act (CCPA; California State Legislature 2018) or the General Data Protection Regulation (GDPR; Council of the European Union 2016). The technical perspective focuses on confidentiality or anonymity, for example, by leveraging technical mechanisms such as encryption, differential privacy, or k-anonymity (Yin et al. 2021). The psychological perspective focuses on privacy perceptions or mental models, for example, by addressing privacy concerns (eg, Smith et al. 1996), privacy awareness (eg, Soumelidou and Tsohou 2021), or privacy literacy (eg, Masur 2020).

Since consumers' privacy expectations are diverse and evolve over time (Bélanger and James 2020), our work is informed by a social perspective on privacy that can account for the diversity and evolving nature of consumers' privacy expectations. We view privacy as a dynamic set of context-dependent expectations that consumers or groups of consumers have with respect to activities such as limiting access to information about oneself, controlling information processing, and expressing one's identity (DeCew 1997). Consumers generally have an understanding of what constitutes appropriate information processing in a particular context and perceive privacy as violated whenever they encounter privacy practices they deem inappropriate (Nissenbaum 2010). Hence, consumers' context-dependent privacy expectations must be accounted for in the design of transparency artifacts.

Extant IS research on transparency typically does not have a strong focus on consumer needs or humanistic goals like privacy. Transparency is generally treated as a strategic issue or an empirical measure for how well system outputs can be explained or predicted. Research approaching transparency as a strategic issue typically focuses on business issues that arise when revealing or concealing information on items or products (Granados et al. 2010)—for example, with respect to sharing of information about transaction details (eg, Nicolaou and McKnight 2006), product prices (eg, Soh et al. 2006), product quality attributes (eg, Liang et al. 2017), or trade-offs between product attributes (eg, Xu et al. 2014). Research employing transparency as a measure of how well system output can be explained generally treats transparency as an empirical outcome in research fields such as government services (eg, Venkatesh et al. 2016), knowledge management (eg, Hornyak et al. 2020), research methods (eg, Paré et al. 2016), or market interactions (eg, Cho et al. 2021). Similarly, extant privacy research typically treats transparency as an antecedent of consumer behaviors (eg, Awad and Krishnan 2006, Betzing et al. 2020, Karwatzki et al. 2017, Martin et al. 2017, Tsai et al. 2011) or as an instrumental design goal (eg, Hosseini et al. 2018, Nussbaumer et al. 2012, Samavi and Consens 2018, Schaub et al. 2017).

Viewing transparency as an antecedent for some desired effect or as an empirical measure is not informative with respect to the prescriptive design focus of TIPP theory. The purpose of TIPP theory is to inform artifact designs useful for establishing TIPP rather than to predict what would happen if IS were more, or less, transparent or how to empirically assess the transparency-related properties of an IS. Hence, we view *TIPP*[5] as a quality of an IS that makes the privacy practices of interest in an IS easy to understand for consumers. With the term *privacy practices*, we refer to organizational privacy practices and not counteractive practices of consumers (Bélanger and Crossler 2011). Accordingly, the term privacy practices captures practices in the IS that are concerned with information collection and use, the protection of information from intrusion, the restriction of access to information, or the facilitation of privacy management (Tavani 2007).

Issues related to technical subsystems of transparency artifacts have been addressed in extant research. Privacy notices have, for instance, been criticized for lacking readability (Milne et al. 2006, Sunyaev et al. 2015), insufficient compliance with fair information practice principles (Rains and Bosch 2009, Schwaig et al. 2006), offering information that is not of interest to consumers (Earp et al. 2005), or lacking fit with the needs of special audiences (Milkaite and Lievens 2020). While these are design shortcomings that must be resolved to make privacy notices more usable, resolving such shortcomings is unlikely to establish TIPP. Looking at the issue from the perspective of privacy as a social value reveals that privacy notices are designed for a social subsystem that is a poor fit for consumers' actual, context-dependent privacy expectations. Put simply, privacy notices are not a useful artifact to reveal the information consumers need to interact with IS in line with their privacy expectations.

A key assumption underlying the design of privacy notices is that "consumers read these [privacy] notices and make decisions according to their overall preferences, including preferences about privacy, price, service offering, and other attributes" (Hoofnagle and Urban 2014, p. 262). That is, instead of adapting technical subsystems (ie, privacy notices) to underserved needs of social subsystems (ie, consumers' privacy expectations), the design of privacy notices is technocentric and requires consumers to adapt to privacy notices. This constitutes a 'dying' relationship between the social and technical subsystems (De

---

[5] Our definition of TIPP is based on the definition of transparency as "the quality of something, such as a situation or an argument, that makes it easy to understand" (Hornby 2000, p. 1383).

Leoz and Petter 2018) because privacy is a pluralistic concept that is always in flux and shaped by continual technological innovation and contextual cues (Mulligan et al. 2016), and privacy notices are too static to adapt to such changes. TIPP theory accounts for this by taking a sociotechnical view on the design of transparency artifacts—in particular, by accounting for privacy as a social value.

## THE DEVELOPMENT OF TIPP THEORY

Similar to Kasper's approach to developing an ISDT to foster decision quality in decision support systems (1996), we employ a conceptual approach for the development of TIPP theory, which is in line with our goal to inform the design of technical subsystems of transparency artifacts by approaching it top-down from a social privacy perspective.

Iivari (2020) identified four different conceptions of ISDT: (1) "theoretical origin of metarequirements and metadesign for the IT artifact" (p. 514), (2) "the relationship between metadesign and metarequirements of the IT artifact" (p. 514), (3) "the relationship between the class artifacts (as defined by metarequirements and metadesign) and the effectiveness of the artifacts" (p. 514), and (4) a union of conceptions 1–3. Since this manuscript is focused on deriving implications from the outer environment (social subsystems of transparency artifacts) for the design of the inner environment (technical subsystems/IT artifact), we use ISDT in the first sense (Walls et al. 1992, 2004) and focus on the "theoretical origin of metarequirements and metadesign for the IT artifact" (Iivari 2020, p. 514). Moreover, we follow Kuechler and Vaishnavi's proposal (2012) of complementing ISDT, which "captures meta-requirements and a meta-design that are applicable to a class of artifacts" (2012, p. 399), with DREPT, which they define as "explanatory/predictive […] theory […] derived from a highly abstract covering theory (kernel theory) that originated in a non-design domain or tacit theory […] [b]ut in which the kernel or tacit theory constructs have been translated into a technology domain" (2012, p. 404).

The development of TIPP theory can be characterized as disciplined imagination, which represents the "view that theory construction involves imagination disciplined by the processes of artificial selection" (Weick 1989, p. 528). Accordingly, we conducted thought trials (Weick 1989) to test potential ideas for resolving the overarching problem of making transparency artifacts useful with respect to consumers' context-dependent privacy expectations. As selection criteria to evaluate the ideas, we focused on *interestingness* and *plausibility* in line with the conceptual nature of our research approach (Weick 1989). To determine whether ideas (or aspects of them) violate the interestingness criterion, we checked whether they were irrelevant, obvious, or absurd (Davis 1971). Ideas that are not helpful for making transparency artifacts useful with respect to consumers' context-dependent privacy expectations were rejected as irrelevant. Ideas that represent general knowledge relevant to the design of transparency artifacts (eg, transparency artifacts should be usable, which is the case for any IS artifact designed for voluntary use) were rejected as obvious. To reject absurd ideas, we checked the pertinent literature for contradicting evidence or conducted our own empirical studies. Furthermore, we grounded ideas in kernel theories to avoid absurd ideas. However, the main purpose of grounding ideas in kernel theories was to establish the plausibility of ideas by demonstrating that the assertions made can be linked, primarily through analogical reasoning, to more general theories.

In the next section, we focus on the final version of TIPP theory. Please refer to the appendix for examples of thought trials and their implications for the development of TIPP theory.

## AN INFORMATION SYSTEMS DESIGN THEORY FOR TRANSPARENCY OF INFORMATION PRIVACY PRACTICES

This section begins with a presentation of our developed DREPT (TIPP Part-1), which focuses on social subsystems of transparency artifacts that are useful with respect to consumers' context-dependent privacy expectations, and concludes with a corresponding ISDT (TIPP Part-2), which outlines a solution space for technical subsystems of transparency artifacts.

### TIPP Part-1: Design-relevant explanatory/predictive theory (DREPT)

Since establishing TIPP necessitates the presentation of information on privacy practices to consumers, the main actors relevant for the emergence of transparency artifacts are consumers as well as the providers of the IS artifacts for which TIPP should be established. Emergence of useful transparency artifacts is expedited in operational settings where both providers and consumers perceive a need for TIPP.

**Consumer perspective.** Since privacy is a pluralistic concept that is always in flux and shaped by continual technological innovation and contextual cues (Mulligan et al. 2016), a proxy concept is necessary to account for privacy in the design of transparency artifacts. In TIPP theory, privacy behaviors are used as a proxy, because they are more directly observable from consumer interactions with IS artifacts than more psychological concepts (eg, privacy concerns, privacy norms, past privacy experiences, privacy awareness) in privacy-related nomological networks (eg, Bélanger and James 2020, Dinev et al. 2015, Li 2011, Smith et al. 2011). We refer to consumer actions that result in information flows or aim to maintain the appropriateness of information processing as privacy behaviors[6] (Nissenbaum 2010).

We propose an initial[7] typology of seven types of consumer privacy behaviors: privacy practice assessment, disclosure, concealment, deletion, information flow management, multiparty privacy coordination, and privacy violation response (Table 1).

*Privacy practice assessment* refers to consumer behaviors that assess the appropriateness of information processing with respect to a consumer's privacy expectations; for example, the reading of privacy notices (Milne and Culnan 2004), the development of a mental model of information processing (Lin et al. 2012), and comparing privacy practices with one's privacy expectations (Wright and Xie 2019).

---

[6] It is important to note that privacy behaviors encompass not only confidentiality-preserving behaviors (Ben-Shahar 2019, Son and Kim 2008). In our globally connected world, with its increasing emergent interactions between technical and social structures in everyday life (Yoo 2010), privacy behaviors aim to maintain the appropriate processing of information, which inevitably also includes behaviors such as the release of information and the management of online identities (Nissenbaum 2010).

[7] The typology captures a wide range of privacy behaviors but is not supposed to be complete because privacy behaviors are bound to change with technological innovation. For example, widespread dissemination of and cheap access to information over the internet made multiparty privacy coordination (Altman 1975) more relevant in IS (Peppet 2011) and some technologies prevent some privacy behaviors by design (eg, blockchains that are designed to prevent anyone from deleting data once it has been entered into the linked list (Kannengießer et al. 2020, Nakamoto 2008)).

*Table 1. Overview of seven different types of privacy behaviors.*

| Privacy Behavior | Description | Examples |
|---|---|---|
| Privacy practice assessment | Determine the appropriateness of information processing with respect to privacy expectations | Develop a mental model of information processing, compare privacy practices with privacy expectations |
| Disclosure | Reveal information to others | Share information, cultivate online identities |
| Concealment | Reveal information to others while distorting, excluding, or hiding information | Use anonymization services, pseudonyms, or encryption |
| Deletion | Erase information after disclosure | Delete information directly, invoke the right to be forgotten |
| Information flow management | Restrict or broaden the flow of released information | Use privacy settings, target information to specific audiences |
| Multiparty privacy coordination | Ensure appropriate use of co-owned information | Anticipate privacy consequences for others, seek approval before disclosure |
| Privacy violation response | React to perceived privacy violations | Spread negative word of mouth, switch to alternative IS |

*Disclosure* refers to consumer behaviors that reveal information to others. Examples of such behaviors include sharing information (Cavusoglu et al. 2016), passively releasing information when using IS (Awad and Krishnan 2006), and cultivating online identities in IS such as social networking services (Wu 2019).

*Concealment* refers to consumer behaviors that reveal information to others while distorting, excluding, or hiding information. Examples of such behaviors include anonymization (Feigenbaum and Ford 2015), the use of multiple identities or pseudonymization (Marwick and boyd 2010), encryption (Whitten and Tygar 1999), and the sharing of distorted or falsified information (Son and Kim 2008).

*Deletion* refers to consumer behaviors that aim to erase information after disclosure. Examples of such behaviors include leveraging user interface features to delete shared information (Young and Quan-Haase 2013) and invoking legal rights, such as the GDPR's 'right to be forgotten' (Council of the European Union 2016), that oblige providers to delete links to information (Newman 2015).

*Information flow management* refers to consumer behaviors that restrict or broaden the flow of information once it has been released. Examples of such behaviors include using opt-in or opt-out features (Cranor 2012), using offered privacy settings (Crossler and Bélanger 2019), and targeting shared information (eg, posts on social media) to make it less discoverable by undesired audiences (Marwick and boyd 2014, Moll et al. 2017).

*Multiparty privacy coordination* refers to consumer behaviors that aim to ensure the appropriate use of co-owned information (Bélanger and James 2020). Examples of such behaviors include anticipating privacy consequences for others, seeking approval prior to disclosure, and identifying privacy norms or policies that are shared with other consumers (Martin 2016, Such and Criado 2018).

*Privacy violation response* is a behavior that is only exhibited after a privacy violation has been perceived by consumers and captures reactions to privacy violations. Examples of such behaviors include initiating privacy-related lawsuits on various grounds—for instance, defamation, discrimination, or the negligence of data protection (Romanosky 2016)—spreading negative word of mouth (Son and Kim 2008), and switching to alternative IS (Martin et al. 2017).

The privacy behavior typology illustrates that privacy behaviors are diverse and versatile, which contributes to the diversity and context-dependency of consumers' privacy information needs since consumers need different information to perform different privacy behaviors (eg, deletion vs. information flow management). Transparency artifacts must reveal the information that consumers need to perform a privacy behavior in order to be useful regarding consumers' context-dependent privacy expectations.

**Provider perspective.** Transparency research in other domains (see Schnackenberg and Tomlinson 2016 for a review), for instance, product transparency in the travel industry (Granados et al. 2010), shows that increasing transparency is a valid strategy for IS providers to increase consumer trust and reap first-mover advantages if they are frontrunners in a market that starts to move toward more transparency (Granados et al. 2010, Schnackenberg and Tomlinson 2016). For establishing TIPP, things are different, because information release cannot solely serve strategy or compliance objectives and must meet social privacy expectations that often exceed or even contradict[8] legal requirements (eg, Milne et al. 2017).

---

[8] Legally mandated cookie disclaimers are, for instance, often perceived as annoying by consumers and fail to incentivize interaction with IS that is aligned with consumers' privacy expectations (Kulyk et al. 2020).

From a privacy perspective, transparency is not "best viewed as a perception of the quality of *intentionally* [emphasis added] shared information from a sender" (Schnackenberg and Tomlinson 2016, p. 1803) since withholding information hinders assessments whether privacy practices meet consumers' privacy expectations (Donaldson and Dunfee 1999). Establishing TIPP may not only be unprofitable but even detrimental for some providers because revealing information on privacy practices always has the potential to cue privacy concerns if consumers do not consider information processing to be appropriate (Nissenbaum 2010). Hence, providers must decide between privacy and secrecy—that is, whether they can retain/attract enough consumers by establishing TIPP or whether the risk that a large proportion of consumers will consider the privacy practices to be inappropriate is too great.

In the following, we refer to IS providers who "view information privacy as 'table stakes'" (Greenaway and Chan 2005, p. 181) and perceive a need for processes to be in place that keep privacy practices secret or obscure (eg, offering only distorted, biased, or opaque information on privacy practices; Granados et al. 2010) as *secrecy providers*. A prominent reason motivating providers to oppose TIPP and act as a secrecy provider is, for instance, what Zuboff termed 'surveillance capitalism' (2015, 2019). On the other hand, we refer to IS providers who can be motivated to establish TIPP as *privacy providers*.[9] Privacy providers have a prosocial stance on privacy, which particularly entails privacy-related organizational actions that go beyond mere compliance with data protection laws, and have a stronger motivation to meet consumers' privacy expectations[10] (Bamberger and Mulligan 2011).

By definition, secrecy providers cannot account for consumers' privacy expectations with respect to TIPP. This constitutes a 'dying' relationship between the social and technical subsystems because one party cannot account for the needs of the other and no transparency artifact useful for establishing TIPP can emerge (De Leoz and Petter 2018).

---

[9] Please refer to Greenaway and Chan (2005) for similar classifications of organizational stances on privacy on a more abstract level (general organizational privacy behaviors) and with a different theoretical foundation that is based on institutional theory (acquiescence strategy *(secrecy provider)* and proactive strategy *(privacy provider)*) and the resource-based view of the firm (customer knowledge capability *(secrecy provider)* and customer relationship capability *(privacy provider)*).

[10] In their study of the privacy practices of 'privacy leaders,' Bamberger and Mulligan ascribe a similar relational stance on privacy to the behavior of privacy providers based on their findings that "[p]rivacy leaders […] emphasized the customer's experience, including 'think[ing] about how this feels from the customer perspective, not what we think the customer needs to know'" (2011, p. 270).

Since potential *secrecy providers* (eg, Google or Facebook) cannot establish TIPP without harming themselves, the alignment of similar services (eg, the search engine DuckDuckGo[11] or the social network federation Fediverse[12]) with consumers' privacy expectations may offer a competitive advantage (Martin and Murphy 2017, Schmidt and Keil 2013) for *privacy providers*. While privacy providers may be interested in using TIPP to gain a competitive advantage by strengthening organizational legitimacy (Bitektine 2011), establishing TIPP is complex (Spiekermann et al. 2019), expends resources with unclear relationships to the core business, and may even reveal trade secrets to competitors (Rudin 2019), thereby reducing potential competitive advantages. Hence, privacy providers may not have sufficient intrinsic motivation to establish TIPP (Smith 1993) and may require extrinsic motivation, in the form of societal pressures (or "external threats"; Smith 1993), to establish TIPP.

The risks of being subjected to the enforcement of data protection laws, such as those of the CCPA (California State Legislature 2018) or the GDPR (Council of the European Union 2016), can motivate privacy providers to establish TIPP (Smith 1993). However, business pressures might make it more profitable for companies to focus on minimal compliance with laws (Bamberger and Mulligan 2011, Greenaway et al. 2015) or even ignore/violate laws (Greenaway et al. 2015, Wall et al. 2016). Societal pressures can also manifest in the form of increasing consumer awareness/demand (Bamberger and Mulligan 2011, Smith 1993) and motivate privacy providers to offer useful transparency artifacts in order to differentiate themselves from competitors (Gerlach et al. 2019, Greenaway et al. 2015). Likewise, societal pressures can manifest in form of competitive pressures if close competitors gain competitive advantages by establishing TIPP (Greenaway and Chan 2005). Privacy providers may further be motivated to leverage transparency artifacts as an information resource to reduce uncertainty about consumers' privacy expectations (Greenaway and Chan 2005) and learn about privacy issues to which data protection laws offer few insights because they cannot account for consumers' evolving and context-dependent privacy expectations (Bamberger and Mulligan 2011, Mulligan et al. 2016).

---

[11] https://duckduckgo.com/about
[12] https://fediverse.party/en/fediverse/

By analyzing the use of transparency artifacts, providers can gain insights into the privacy practices that are of particular interest to consumers. This is helpful information for providers seeking to fulfill their "positive duty to identify and respect privacy expectations of users" (Martin 2020, p. 88) and to ascertain where changes to privacy practices are warranted or promising so that they are perceived as more appropriate by consumers. Likewise, such information can be helpful for providers to understand when privacy practices must be governed with particular care versus when privacy practices can freely emerge, based on consumers' interest (or lack of interest) in privacy practices in various contexts. Table 2 gives an overview of the diverse ways in which societal pressures can manifest to motivate privacy providers to establish TIPP.

In a nutshell, whether providers establish TIPP depends on business considerations and is a decision based on rational organizational behavior (Ben-Shahar 2019, Day and Stemler 2019, Smith 1993). Since the legal requirements of TIPP can already be satisfied through superficial approaches, such as posting privacy notices that do not create value for and are of little interest to consumers (Obar and Oeldorf-Hirsch 2020), the emergence of useful transparency artifacts will be expedited in operational settings characterized by complementary societal pressures that encourage providers to take establishing TIPP seriously. Positive consumer experiences with transparency artifacts in one domain can raise consumer demand for transparency artifacts in other domains. This can make transparency artifacts worthwhile for privacy providers due to the increasing societal demands they face and their ability to learn from the successes and failures of previously tested approaches (Bamberger and Mulligan 2011).

**The emergence of transparency artifacts.** Since TIPP theory is supposed to account for a social value (privacy) by fostering the emergence of transparency artifacts in line with the needs of individual actors (consumers and providers), we consolidate the relevant mechanisms in the outer environment of transparency artifacts in Figure 1 in analogy[13] to Coleman's macro-micro-macro model (1986). The macrolevel refers to the societal level and the microlevel refers to the individual level of analysis (Markus and Robey 1988). Coleman's macro-micro-macro model captures how situations at the societal level motivate actions at the individual level and how these actions combine to produce the envisioned desirable social outcome at the macrolevel (1986).

*Table 2. Overview of potential external factors that can, in a complementary fashion, increase societal pressures on IS providers to motivate them to establish TIPP.*

| Factor | Description | Example |
|---|---|---|
| Enforcement of data protection laws | Providers that face a high likelihood of being subjected to legal enforcement suits they are likely to lose (eg, providers that are publicly funded, have large user bases, or have to access sensitive information) can be motivated to establish TIPP for legal-compliance reasons. | Many companies started to offer or reworked their privacy notices around the time the GDPR came into force (Degeling et al. 2019). |
| Consumer awareness/ demand for TIPP | Providers, where profit generation depends on continuous direct-to-consumer transactions, can be motivated to establish TIPP by increases in consumer awareness of and demand for TIPP (eg, if privacy practices raise negative publicity). | After negative publicity about the unconsented-to use of Uber's 'God View' tool by employees to track the movement of individuals (eg, critical journalists), Uber started to publish a privacy notice forbidding such unconsented-to practices in 2014 (Bhuiyan and Warzel 2014). |
| Competitive advantages due to TIPP | Providers that can offer similar service quality with privacy practices that are perceived as more appropriate by consumers than the privacy practices in rival IS can leverage TIPP as a competitive advantage. | To differentiate from other mainstream mobile operating systems, Apple introduced a feature, called 'App Privacy Report' in iOS 15.2, which presents information on permissions used and third parties contacted by apps running on a user's mobile device (Cipriani 2021). |
| Competitive pressures due to TIPP | Providers faced with competitors gaining competitive advantages by establishing TIPP can be motivated to establish TIPP themselves to preserve/increase their market share. | Many online advertisers adopted the YourAdChoices icon, which associates online advertisements (ads) with a link offering information why the ad was placed (Digital Advertising Alliance 2022). |
| Uncertain privacy expectations | Providers aiming to improve or monitor the alignment of privacy practices with consumers' privacy expectations can be motivated to provide transparency artifacts as resources to survey and better understand consumers' privacy expectations. | Wikimedia operates multiple channels to solicit suggestions for improving its privacy notice and stipulates an open comment period of at least 30 days to solicit user feedback before making any substantive changes to its privacy notice (Wikimedia Foundation 2021). |

---

[13] Coleman's macro-micro-macro model stems from sociology, which aims to understand how things are in society. Hence, we use Coleman's model as an analogy since the purpose of TIPP theory is to explain and prescribe what should be built to establish TIPP and not to describe the current state of affairs with respect to TIPP.

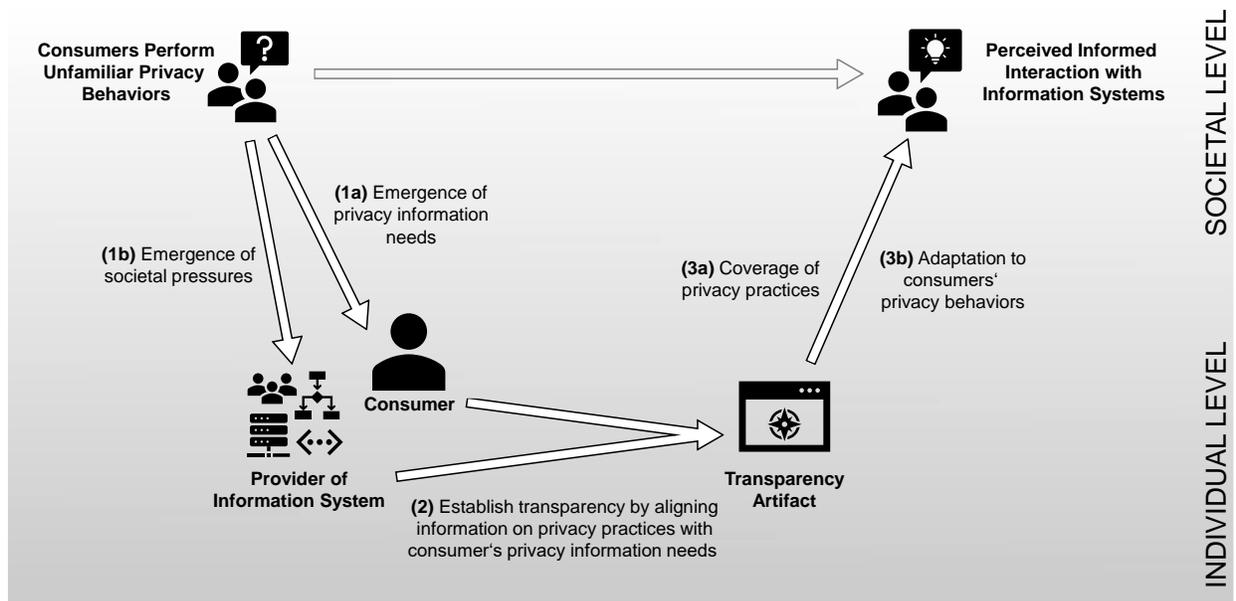

*Figure 1. Macro-micro-macro representation of the DREPT (design-relevant explanatory/predictive theory) on how the outer environment of transparency artifacts constrains and promotes the emergence of useful transparency artifacts that correspond to consumers' evolving and context-dependent privacy expectations.*

The macrolevel transformation brought about by the emergence of useful transparency artifacts constitutes a transformation from a state without TIPP, where consumers perform unfamiliar[14] privacy behaviors, to a state with TIPP, where consumers perceive their interaction with IS to be informed. Once consumers need more information to perform privacy behaviors when interacting with an IS and the provider is motivated to reveal the needed information to consumers, useful transparency artifacts can emerge at the microlevel (situational mechanism; Hedström and Swedberg 1996). The purpose of transparency artifacts is to establish TIPP by offering the information on privacy practices that consumers need to perform privacy behaviors (individual action mechanism; Hedström and Swedberg 1996). In the following discussion of the ISDT-part of TIPP theory, we present a corresponding solution space for technical subsystems of transparency artifacts that outlines artifact designs useful for reaching a macrolevel state where consumers perceive their interaction with IS to be informed (transformational mechanism; Hedström and Swedberg 1996).

---

[14] With unfamiliar privacy behaviors, we refer to a situation where consumers perform privacy behaviors (eg, disclosure) while not being able to understand what they are doing because they do not have access to information on the corresponding privacy practices. An example is, for instance, the Cambridge Analytica incident where many Facebook users inadvertently shared personal information of their social media contacts, which ultimately led to a huge dataset that could have been used for psychographic profiling to manipulate public opinions on national levels (Hu 2020, Sunyaev 2020).

In brief, our presentation of the developed DREPT offers an explanation how the outer environment of transparency artifacts constrains and promotes the emergence of transparency artifacts that are aligned with consumers' context-dependent privacy expectations. In other words, we explain how transparency artifacts can contribute to a transition from a socially undesirable state without TIPP, where consumers are subjected to the risks associated with performing unfamiliar privacy behaviors with unknown consequences (Sinnreich and Gilbert 2019), to a state with TIPP, where consumers' information needs are satisfied and they perceive their interaction with IS to be informed.

**TIPP Part-2: Information systems design theory (ISDT)**

In this section, we capture the resulting design-product implications for technical subsystems of transparency artifacts in line with Walls et al.'s ISDT conceptualization (1992, 2004). This section is organized according to the logical progression of Walls et al.'s ISDT conceptualization and introduces kernel theories, discusses metarequirements, presents a corresponding metadesign, and concludes by summarizing the fit between the metarequirements and metadesign.

**Kernel theories.** TIPP theory is grounded in two kernel theories from the domains of business ethics and educational psychology—integrative social contracts theory and cognitive load theory. Integrative social contracts theory yields insights how to resolve the dichotomy of the privacy-related social contracts underlying TIPP between providers and consumers. In addition, cognitive load theory describes how situations in which transparency artifacts are misaligned with the cognitive capacities of consumers can be avoided.

*Integrative social contracts theory* aims to bridge the gap between the universal norms guiding human behavior and the diverse explicit or implicit social norms valued in contextualized communities. A community is "a self-defined, self-circumscribed group of people who interact in the context of shared tasks, values, or goals and who are capable of establishing norms of ethical behavior for themselves" (Donaldson and Dunfee 1994, p. 262). Regarding TIPP theory, relevant communities are groups of consumers who exhibit shared tasks, values, or goals. Given the widespread dissemination and the diverse affordances of IS (Yoo 2010), consumers belong to multiple communities; the *relevant* community is determined by a consumer's current context.

Integrative social contracts theory is based on two types of social contracts. First, macrosocial contracts are hypothetical, normative contracts that govern general economic behavior. Macrosocial contracts specify the rules that all members of a society would agree upon "when asked what rules they would want applied to them in the context of economic transactions, under the condition that they do not know the position they would occupy under the rules" (Donaldson and Dunfee 1995, p. 93). Regarding TIPP theory, laws, and especially data protection laws (Greenleaf 2014), can be considered macrosocial contracts (Donaldson and Dunfee 1999) because they represent the outcome of a democratic consensus-finding process to establish "the basic rules without which an ordered society is impossible" (Fuller 1969, p. 5).

Second, microsocial contracts are implicit contracts representing social norms valued by specific communities and practiced in the real world. Microsocial contracts account for contextual influences by enabling actors to develop their own rules governing behavior in distinct communities. An example of two communities with similar privacy behaviors but different privacy-related social norms is given by *whistleblowers* (Elliston 1982) and *social media influencers* (Lou and Yuan 2019). Both communities perform privacy behaviors to convincingly share information (ie, disclosure), but they need different information on privacy practices. Whistleblowers need information on privacy practices that enable them to conceal unnecessary details (eg, current location or real identity) to protect themselves, while influencers need information on privacy practices that enable them to represent themselves as authentically as possible to establish rapport with their followers.

Furthermore, integrative social contracts theory distinguishes between two types of norms for microsocial contracts—authentic norms and obligatory norms. Authentic norms fulfill basic conditions to constitute ethical norms; that is, they are "grounded in informed consent and buttressed by a right of exit" (Donaldson and Dunfee 1994, p. 262). Although informed consent does not have to be explicit, most members of contextualized communities must approve of norms, disapprove of deviances from norms, and act in accordance with norms. Authentic norms are also obligatory norms if they fulfill the additional condition of being compatible with *hypernorms*. Hypernorms "entail principles so fundamental to human existence that they serve as a guide in evaluating lower level moral norms" (Donaldson and Dunfee 1994, p. 265). Hypernorms are norms that are valued across cultures and ensure that macrosocial contracts do not sanction arbitrary microsocial contracts. With respect to TIPP, the *notice* privacy principle[15] represents a hypernorm (Cranor 2012).

Obligatory norms represent the contact point that establishes the link between the privacy expectations of consumer communities and the privacy practices in an IS, which are usually aligned with macrosocial contracts. Establishing TIPP obliges providers to align privacy practices in the IS not only with macrosocial contracts but also with the various obligatory norms upheld by the consumer communities the IS is supposed to serve. That is, providers must go beyond mere compliance with data protection laws and account also for consumers' privacy expectations in the design of useful transparency artifacts. Otherwise, transparency will not be established because consumers will most likely not be offered the information they need to interact with IS in line with their privacy expectations.

*Cognitive load theory* is concerned with fostering understanding and learning by deriving implications for instructional design based on a model of human cognitive architecture (Paas and Ayres 2014, Sweller et al. 1998). Cognitive load theory is based on a model of human cognitive architecture comprising constrained working memory and unlimited long-term memory. All understanding and learning occurs in the working memory, which can handle only a small number of information elements (Paas and Ayres 2014). Once novel knowledge is understood and learned, it is stored in long-term memory, which can store an unlimited amount of knowledge with an arbitrary level of complexity (Sweller et al. 1998). Long-term memory allows humans to perform complex information acquisition tasks because recalled knowledge only consumes a single element of working memory capacity and thus frees up cognitive resources. Cognitive resources are consumed by two types of cognitive load, intrinsic and extraneous load (Kalyuga 2011). Intrinsic load is determined by the number and interactions of elements relevant to information acquisition tasks and individual expertise (van Merriënboer and Sweller 2005). Extraneous load constitutes noise irrelevant to the tasks at hand and impedes understanding by wasting cognitive resources.

---

[15] The notice privacy principle is a combination of the openness and disclosure fair information practice principles and postulates that consumer information is not processed in secret and that consumers can find out what information is collected about them and how it is used (US Federal Department of Health Education and Welfare 1973).

Cognitive load theory is a useful kernel theory for TIPP theory because consumers cannot process information offered on privacy practices when their cognitive capacities are overstrained (Alashoor et al. 2022). Cognitive load theory complements integrative social contracts theory by explaining how consumers' information needs and constraints of cognitive resources can be met. Effective designs for fostering understanding focus on reducing extraneous load and maintaining intrinsic load at levels that harness working memory capacity but do not overload it (van Merriënboer and Sweller 2010). That is, useful transparency artifacts should reveal only the information consumers need to interact with IS in line with their privacy expectations. Cognitive load theory substantiates the claim that transparency artifacts must adapt to the tasks consumers wish to perform, must communicate information pertaining to consumers' current tasks, and must be able to account for variability in consumers' tasks over time and across individuals (Rouse and Rouse 1984).

**Metarequirements.** Insights from the two kernel theories and the DREPT counterpart presented in the previous section constitute the foundation for the two metarequirements of a new generation of transparency artifacts based on TIPP theory: coverage and adaptivity. As we elaborate in the following, transparency artifacts are deemed useful for establishing TIPP if they fulfill the coverage and the adaptivity metarequirements.

"Coverage refers to the comprehensiveness or depth of the information provided" (Metzger 2007, p. 2079) by a transparency artifact. Since TIPP is a quality of an IS that makes the privacy practices easy to understand for consumers, establishing TIPP necessitates access to information on the privacy practices in the IS. Establishing coverage constitutes a challenge because the information that consumers consider to be relevant evolves over time and cannot be prespecified (Turner and Dasgupta 2003, Yun et al. 2019). Moreover, as also supported by integrative social contracts theory, individual consumers have different privacy expectations depending on their current community or context (Donaldson and Dunfee 1999, Mulligan et al. 2016, Nissenbaum 2010). Consumers' evolving and context-dependent privacy expectations lead to diverse privacy information needs, which must be met to establish TIPP; otherwise, TIPP would only be established for some consumers in certain contexts. For providers of transparency artifacts, this entails that they must maintain an evolving documentation of privacy practices, which yields the information of interest to the diverse consumer communities served by the IS. Hence, the coverage metarequirement:

> ***Coverage***: *Establishing transparency of information privacy practices requires maintenance of the information necessary to satisfy the evolving privacy information needs of consumers.*

Establishing coverage is a necessary but not sufficient metarequirement for establishing TIPP. As demonstrated by the practice of posting privacy notices (McDonald and Cranor 2008, Sunyaev et al. 2015), presenting consumers with a lot of information leads to information overload and prevents consumers from retrieving information of interest to them (McDonald and Cranor 2008, Milne and Culnan 2004, Sheng and Simpson 2014). Consequently, establishing TIPP also requires adaptivity. Transparency artifacts are adaptive if they are "able to change when necessary in order to deal with different situations" (Hornby 2000, p. 14). This means that transparency artifacts should not only offer the information of interest to different communities (Donaldson and Dunfee 1999), but they must also feature communication interfaces that are adaptive to the different information needs of the different communities served by the IS. To satisfy information needs, consumers shift among information-seeking strategies (eg, searching for information, acquiring information, comparing information) until they have fulfilled or abandoned their search goals (Xie 2000). Accordingly, establishing TIPP necessitates that transparency artifacts are adaptive to consumers' different information-seeking strategies so that changes in consumers' information needs across contexts and different consumer communities can be accounted for (Rouse and Rouse 1984). Cognitive load theory supports the adaptivity metarequirement by positing that effective artifact designs should lead to levels of intrinsic load that do not overload working memory capacity (van Merriënboer and Sweller 2010). Accordingly, information presentation must be adaptive to levels of intrinsic load that consumers can handle. Hence, the adaptivity metarequirement:

> *Adaptivity*: *Establishing transparency of information privacy practices requires adaptation of information presentation to deviations in consumers' privacy information needs.*

Neither coverage nor adaptivity is a metarequirement that is sufficient to establish TIPP on its own. Focusing on coverage will lead to information overload and thereby prevent information retrieval, given the demands of everyday life. Focusing on adaptivity will lead to good alignment with the demands of everyday life but will only partially address the information needs of the consumers the IS serves. To actually be useful, transparency artifacts must achieve both metarequirements (coverage and adaptivity) in order to reveal the information of interest to consumers while allowing for context-dependent adaptations.

**Metadesign.** To achieve the coverage and adaptivity metarequirements, transparency artifacts must offer the information on privacy practices that consumers need for performing privacy behaviors. In terms of integrative social contracts theory (Donaldson and Dunfee 1999), transparency artifacts support the alignment between the macrosocial contracts governing privacy practices in the IS with the microsocial contracts of the consumer groups served by the IS. In essence, this means that two negative feedback loops[16] (Ramaprasad 1983) that control the achievement of both the coverage and the adaptivity metarequirement—that is, the *capacity* of transparency artifacts to establish TIPP—must be implemented (Figure 2).

---

[16] A feedback loop measures the divergence between a desired level and an actual level of something and attempts to adjust it. Positive feedback loops widen the gap while negative feedback loops close the gap (Ramaprasad 1983).

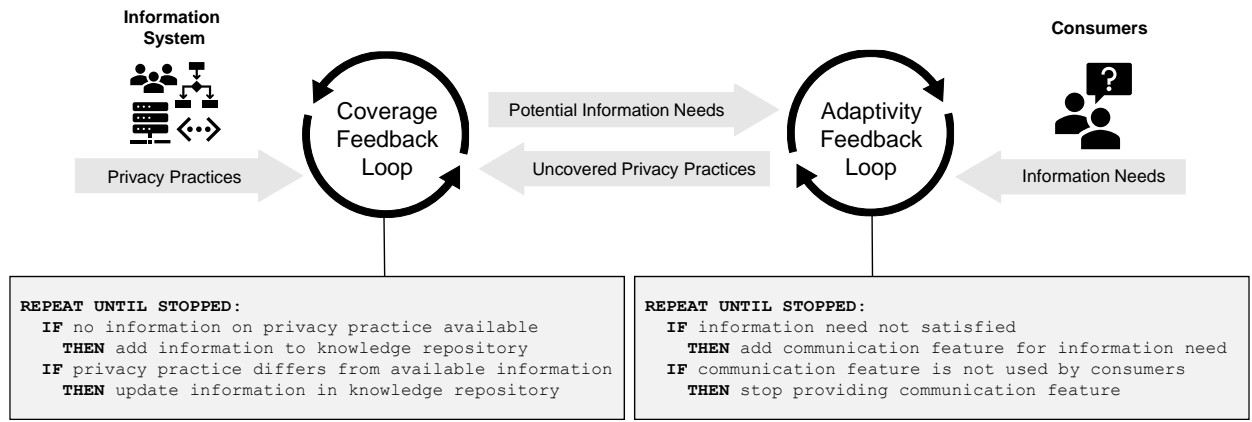

*Figure 2. The two interacting negative feedback loops represent the metadesign of the inner environment (technical subsystems) of transparency artifacts. The arrows represent inputs for the feedback loops exchanged within the technical subsystems and retrieved from the outer environment (social subsystems) of transparency artifacts.*

In the coverage feedback loop the privacy practices are used as input and compared with the privacy practices on which information is available in transparency artifacts. The absence of such information in a transparency artifact triggers corresponding actions that contribute the missing information. Conversely, actions that update information in transparency artifacts are triggered, if the information does not correspond with actual privacy practices (eg, due to a planned feature update, workarounds of employees not following corporate policies, or a successful IS security attack).

In the adaptivity feedback loop consumers' information needs are used as input and compared with information needs that can be satisfied by transparency artifacts. If such information needs cannot be satisfied, actions are triggered that add the missing features to the transparency artifact. Conversely, actions that remove features from transparency artifacts are triggered, if consumers stop exhibiting corresponding information needs.

In concert, the coverage and adaptivity feedback loops serve two main purposes. The coverage feedback loop ensures that information on privacy practices is available in transparency artifacts; the adaptivity feedback loop ensures that features useful for communicating information on privacy practices to consumers are available in transparency artifacts so that consumers' information needs can be satisfied. The value of a transparency artifact designed according to TIPP theory is greater than the value of its parts since additional positive effects can be achieved through joint consideration of both feedback loops.

For instance, consumers may presume negative provider intentions if privacy practices remain unknown (Oulasvirta et al. 2014). This constitutes a problem because the coverage feedback loop only promotes presentation of information on privacy practices carried out but not information on other privacy practices of interest to consumers. Such issues can be resolved by using the privacy practices of interest detected in the adaptivity feedback loop as an additional input for the coverage feedback loop. Another challenge to establishing TIPP is that consumers' privacy information needs arise from anomalous states of knowledge where, "in general, the user is unable to specify precisely what is needed to resolve that anomaly" (Belkin et al. 1982, p. 62). Consequently, consumers' information needs that are used as input for the adaptivity feedback loop may not reveal all the information on privacy practices needed to satisfy their information needs. Such shortcomings can be resolved by feeding the information on privacy practices detected in the coverage feedback loop as potential information needs into the adaptivity feedback loop so that corresponding features can be added to transparency artifacts before consumers demand it. This endows transparency artifacts with a proactive capacity to meet future consumer information needs.

**Fit between metarequirements and metadesign.** The inner environment of transparency artifacts produces output based on the integration of information on privacy practices with information about consumers' privacy information needs and is thus aligned with the DREPT-part of TIPP theory. The two interacting negative feedback loops allow for adaptation to changes in privacy practices or consumers' privacy expectations. Through transformation of the input of one feedback loop so that it can also be used as input for the other, and vice versa, additional desirable outcomes can be achieved, which cannot be achieved by either the coverage or the adaptivity feedback loop on its own. Conceptualizing IS design to establish TIPP in such a way not only allows for transparency artifacts that are adaptive to the contextual demands of consumers' everyday lives (Mulligan et al. 2016, Nissenbaum 2010, Yoo 2010) but also serves as a new source of organizational learning (Maitlis and Christianson 2014) to better tailor IS to consumers' context-dependent privacy expectations (Culnan 2019, Milne et al. 2017). In other words, building transparency artifacts based on TIPP theory serves not only consumer needs, by revealing the information consumers need to interact with IS in line with their privacy expectations, but is also helpful for providers to better tailor privacy practices to consumers' privacy expectations, thereby, making IS more appealing by reducing the likelihood for inadvertent privacy violations.

## DISCUSSION

**Implications for research on the design of transparency artifacts**

TIPP theory establishes a bridge from the complexity of the privacy concept to a metadesign for transparency artifacts that is useful for establishing TIPP in any IS. This lays the theoretical groundwork for informing future artifact designs that establish transparency by revealing the information consumers need to interact with IS in line with their privacy expectations.

TIPP theory extends prior research on the design of transparency artifacts: In contrast to prior research, it is not constrained by the narrow assumption that consumers' privacy behaviors result from a privacy calculus. TIPP theory is useful for informing the design of transparency artifacts that can establish TIPP for consumers performing a wide range of privacy behaviors. The design of useful transparency artifacts must accomplish more than making comprehensive sets of information available to consumers. To be aligned with consumers' context-dependent information needs and avoid information overload, transparency artifacts useful for establishing TIPP must also be adaptive and reveal the information on privacy practices that consumers need in their current context. Otherwise, consumers will either not be offered the information they need to interact with IS in line with their privacy expectations or be unable to digest the information of interest due to information overload.

To serve as a theoretical foundation for artifact designs useful for establishing TIPP, TIPP theory comprises a DREPT- and an ISDT- part. The DREPT-part focuses on relevant mechanisms in social subsystems of transparency artifacts and explains when artifacts useful for establishing TIPP can emerge: (1) For consumers to have a reason to use transparency artifacts, they must have a need for information to perform privacy behaviors. (2) To motivate providers to satisfy consumers' information needs, sufficient societal pressures must be present so that information on privacy practices offered satisfies consumers' information needs.

The ISDT-part of TIPP theory translates the explanatory knowledge contained in the DREPT-part, which is grounded in more abstract privacy research, into a metadesign for technical subsystems of transparency artifacts, which is grounded in kernel theories. We used integrative social contracts theory as kernel theory to clarify the interplay between the hypernorms (eg, the *notice* privacy principle) and macrosocial norms (eg, data protection laws) that guide societal behavior on a general level and the diverse microsocial norms cherished by contextualized communities in the real world. Cognitive load theory complements the insights offered by integrative social contracts theory by shedding light on how to avoid situations where transparency artifacts are misaligned with the cognitive capacities of consumers.

We concluded our explication of the ISDT-part of TIPP theory with an outline of an abstract, technology-agnostic metadesign of transparency artifacts useful for establishing TIPP that consists of two interacting feedback loops (adaptivity and coverage feedback loop) that should be implemented in a way that fits the operational setting of the IS for which TIPP should be established. In short, the DREPT-part of TIPP theory focuses on social subsystems of transparency artifacts to clarify the key mechanisms that have to be present so that useful artifacts can emerge, whereas the ISDT-part focuses on corresponding, prescriptive design knowledge for technical subsystems. Both parts of TIPP theory should be taken into account for the emergence of useful transparency artifacts.

Beyond linking the design of transparency artifacts to privacy as a social value, so that transparency artifacts can better meet consumers' context-dependent information needs, TIPP theory departs from extant transparency-focused privacy research in three main ways. First, the previous literature has conceptualized transparency-related constructs as an antecedent of privacy behaviors (eg, Awad and Krishnan 2006, Betzing et al. 2020, Karwatzki et al. 2017, Martin et al. 2017, Tsai et al. 2011). TIPP theory shows that privacy behaviors can also inform the design of artifacts because they are observable from user interactions with an IS and are thus useful for identifying the information needs that are hopefully served by transparency artifacts. Taking privacy behaviors into account when designing transparency artifacts, will allow for building artifacts that can establish TIPP, even in situations where consumers need information but are not inclined to perform a privacy calculus (Dinev et al. 2015), for instance, when they are cognitively depleted or in a positive mood (Alashoor et al. 2022). The privacy behavior typology (Table 1) illustrates the diversity of privacy behaviors that should be accounted for in the design of transparency artifacts. When instantiating TIPP theory, the adaptivity feedback loop must be designed in a way that results in useful interfaces to present the information relevant to the contexts in which the IS is commonly used.

Second, provider attention to privacy has been conceptualized as a spectrum of feasible strategies resulting from a trade-off between the provider's desires to access consumer information and compliance with data protection laws or consumers' privacy expectations (eg, Culnan 2019, Feigenbaum et al. 2002, Gal-Or et al. 2018, Gerlach et al. 2019, Greenaway et al. 2015, Greenaway and Chan 2005, Wall et al. 2016). Regarding TIPP, the spectrum of feasible strategies boils down to a dichotomous decision. A trade-off between multiple feasible strategies would be unethical because, according to integrative social contracts theory, the key requirements for ethical norms are that most members of contextualized communities must approve of the norm, disapprove of deviance from the norm, and act in accordance with the norm (Donaldson and Dunfee 1994). Secrecy providers that act against data protection laws (ie, macrosocial contracts) or consumers' privacy expectations (ie, microsocial contracts), acting instead in accordance with their own objectives by withholding information, violate such key requirements for ethical norms. By failing to account for consumers' context-dependent information needs and privacy expectations by design, secrecy providers will be unable to offer useful transparency artifacts. This situation is similar to greenwashing, where companies misrepresent their ecological impact in a more positive light (Marquis et al. 2016), which can lead to reputational harm for companies, when consumers realize that actual ecological practices do not correspond to the communicated practices (Nyilasy et al. 2014). Secrecy providers may face similar negative consequences once privacy practices perceived as inappropriate are revealed; thus, they are well advised to keep privacy practices secret or obscure if possible. However, for privacy providers, who are willing to align information processing with consumers' privacy expectations, it makes sense to reveal information of interest to consumers. The metadesign presented in TIPP theory (Figure 2) can serve as an abstract blueprint for such improvements.

Third, privacy notices are seen as "the primary information source for individuals to evaluate marketer privacy practices before disclosing information" (Slepchuk and Milne 2020, p. 90). Many proposals seek to improve privacy notices—for instance, with respect to their timing, communication channel, or modality (eg, Schaub et al. 2017), by converting them into personalized indicators (eg, Tsai et al. 2011), or by summarizing them (eg, Zaeem et al. 2018). TIPP theory sheds some light on why improving privacy notices does not go far enough to establish TIPP. The improvement of privacy notices aligns well with the general idea of cognitive load theory (Kalyuga 2011, Sweller 1988, Sweller et al. 2019) but does not account for the insights revealed by integrative social contracts theory (Donaldson and Dunfee 1999). Privacy notices are predominantly legal documents that can convey compliance with macrosocial contracts, such as data protection laws or fair information practice principles (Greenleaf 2014, Milne and Culnan 2002). However, as revealed by Milne et al. (2017), consumers' privacy risk perceptions are inconsistent—an amalgamation of physical, monetary, psychological, and social risk perceptions—and thus differ from the assumptions about information sensitivity underlying data protection laws such as the GDPR (Council of the European Union 2016). That is, privacy notices do not maintain alignment with the diverse and evolving microsocial norms of the communities for which TIPP should be established (Bélanger and James 2020, Donaldson and Dunfee 1999). In other words, privacy notices are not a useful transparency artifact because they overload consumers with information irrelevant to their current context. Different consumers perceive privacy risks differently (Milne et al. 2017, Mulligan et al. 2016) and the resulting information needs are context-dependent (Nissenbaum 2010). Hence, the key design rationale for building transparency artifacts based on a coverage feedback loop interacting with an adaptivity feedback loop is that this would result in solutions for establishing TIPP that can dynamically adapt to the evolving privacy practices in an IS and the divergent and context-dependent privacy information needs of consumers.

**Guidance for practitioners aiming to instantiate transparency artifacts**

For practitioners, the main implication of TIPP theory is that they must do more than posting privacy notices, if they actually want to enable consumers to interact with IS in line with their privacy expectations and to reduce the likelihood for negative repercussions due to inadvertent privacy violations. Practitioners who see value in establishing TIPP and aligning their privacy practices with consumers' privacy expectations should instantiate transparency artifacts that implement both adaptivity and coverage feedback loops. However, practitioners do not have to start from scratch and can leverage existing guidance and tools to instantiate useful transparency artifacts. To account for the nature of privacy, which is a concept that is in constant flux (Mulligan et al. 2016), efforts suitable for establishing TIPP can be outlined as iterative approaches with at least three steps: (1) a privacy impact assessment, (2) coverage control, and (3) adaptivity control.

A *privacy impact assessment* is "a systematic risk assessment that scrutinises the privacy implications of [a company's] operations and personal data handling practices" (Oetzel and Spiekermann 2014, p. 126). Guidelines for privacy impact assessments have been developed since the 1990s and differ in terms of their scope and level of detail (see Clarke 2009 for a review). Some guidelines focus on lists of questions to support risk assessment (eg, Henriksen-Bulmer et al. 2019, Mantelero 2018) or business processes and documents to be produced (eg, Oetzel and Spiekermann 2014), while others focus on identifying potentials for privacy risk reduction (eg, Senarath and Arachchilage 2019) or on numerical quantifications of risks (eg, Alemany et al. 2018, Hart et al. 2020).

For the design of transparency artifacts, privacy impact assessments are helpful for gaining an overview of the privacy practices in the IS for which TIPP should be established. One focal activity is collecting system documentation to identify the privacy practices in the IS—in particular, what information is collected and how it is used and shared. Afterward, privacy practices can be mapped to privacy risk perceptions common to the IS—gleaned, for example, from privacy concern surveys (eg, Jin et al. 2021). Based on this mapping of privacy practices to privacy risk perceptions, a set of privacy practices that are of particular relevance for establishing TIPP can be identified (ie, privacy practices that also trigger high privacy risk perceptions). These privacy practices can serve as an initial set of privacy practices for which information must be available in transparency artifacts.

During operation, the specific information that needs to be available in transparency artifacts is managed in the *coverage control step*, which comprises at least two parts (see Figure 2): (1) privacy practice cataloging and (2) privacy practice monitoring. The purpose of privacy practice cataloging is to maintain an overview of privacy practices in the IS. The purpose of privacy practice monitoring is to test whether the actual privacy practices violate the intended, catalogued privacy practices. Together they can control fulfillment of the coverage metarequirement by identifying mismatches between intended and actual privacy practices (ie, demand for action by the IS provider).

For IS where privacy practices are easy to assess (eg, IS where information is only processed on local devices), a written document or list may be sufficient for *privacy practice cataloging*. Machine-interpretable solutions are generally preferable in IS where privacy practices are subject to frequent changes, numerous, and harder to track (eg, in applications running on distributed ledger technologies Sunyaev et al. 2021). The Platform for Privacy Preferences Project (P3P; Reagle and Cranor 1999) offers a domain-specific language for encoding privacy practices. However, since P3P was often incorrectly used and had issues with expressiveness (Lämmel and Pek 2013), it was ultimately retired (Cranor et al. 2018). As an alternative to P3P, access control languages, such as the eXtensible Access Control Markup Language (XACML; Anderson 2006), could serve as a foundation for encoding privacy practices. However, providers could also develop their own encoding in JavaScript Object Notation (JSON) or the Extensible Markup Language (XML; Severance 2012) in order to be as flexible as possible and to gain the capacity to document uncommon or complicated privacy practices (Bartsch et al. 2022).

Audits of standard operating procedures and source code reviews represent an intuitive solution for *privacy practice monitoring*. However, such approaches only reveal the discrepancies between cataloged and intended privacy practices. Live information on actual privacy practices could be obtained by extracting information from security technologies that the provider may already have in place. Approaches for static or dynamic code analysis can, for example, be used to automate the identification of mismatches between intended privacy practices and the source code in use (eg, Brüggemann et al. 2019, Yu et al. 2018). Intrusion detection systems are helpful for identifying unknown or undesired privacy practices (Axelsson 2000). Protocols for penetration tests (Bishop 2007) can also be extended to gain insights into how external actors can disrupt the fit between intended and actual privacy practices.

Once information on privacy practices is available in transparency artifacts, it can be communicated to consumers in the *adaptivity control step*, which also comprises at least two parts (see Figure 2): (1) information needs detection and (2) a communication feature provision. The purpose of information needs detection is to discover what information is relevant to consumers in their current context. The purpose of the communication feature provision is to display information on privacy practices in a way that can satisfy consumers' current information needs. Together, they can control the fulfillment of the adaptivity metarequirement by adapting the presentation of information on privacy practices to consumers' information needs.

A simple approach to *information needs detection* is to allow consumers to select corresponding communication features from an application menu. Depending on the complexity of the IS and its use cases, this may, however, lead to a long list of communication features, which may result in too much effort for consumers trying to identify a suitable communication feature. Conversational agents trained for privacy questions (eg, PriBot; Harkous et al. 2018) could serve as an interactive interface to detect information needs. However, the high degree of automation would make it difficult to ensure that consumers are being guided in the direction they desire; further, conversational agents may themselves raise additional privacy concerns (Rajaobelina et al. 2021). Moreover, training the language model and keeping it up to date for their own IS may exceed the expertise and resources of some providers. Nevertheless, conversational agents present an interesting option in cases where voice-based user interfaces are employed in the IS anyway. Similar to the transfer of consumers' privacy settings between domains (eg, Raber and Krüger 2022, Shanmugarasa et al. 2022), consumers' information needs could be inferred based on information needs they have already exhibited in similar IS and contexts. However, unresolved challenges include how to avoid the introduction of additional privacy risks and concerns resulting from the necessary exchanges of user models across different IS and how to detect and account for differences between contexts (Raber and Krüger 2022). As discussed by Rubinstein and Good (2013), a more effortful but also more thorough approach would be to leverage user experience design methods to better understand consumers' information needs and adapt transparency artifacts accordingly. User studies could reveal what information needs consumers exhibit in which contexts and corresponding communication features could be made available. However, such studies would be periodically necessary across the entire lifecycle of the IS, due to the dynamic nature of privacy (Mulligan et al. 2016, Rubinstein and Good 2013). On the upside, providers could leverage the study protocols and findings for evidence-based demonstrations on the usefulness of their transparency artifacts.

Design principles for effective *communication features* have been comprehensively reviewed by Schaub et al. (2015, 2017), who distilled them into a design space with four dimensions: timing, channel, modality, and control. The dimension timing demands attention when to display what information (eg, just-in-time or persistently). The dimension channel concerns where to display the information (eg, in a companion website for a smartwatch with limited display size). The dimension modality focuses on how to communicate information (eg, as text, with icons, or with sound). Finally, the dimension control determines how disruptively information is communicated (eg, forcing user interaction or on demand).

An overview of the main steps in iterative approaches for the instantiation of transparency artifacts is presented in Figure 3. The good news is that practitioners can draw from extant guidance on how to instantiate useful transparency artifacts. Yet, there is no one-size-fits-all solution for how to instantiate transparency artifacts. How to best instantiate transparency artifacts depends on the expertise of and resources available to the IS provider and the complexity of privacy practices which need to be made transparent. As a general rule, providers should strive to reduce the complexity of information processing and of privacy practices; this will make it easier to reveal the information consumers need to interact with IS in line with their privacy expectations.

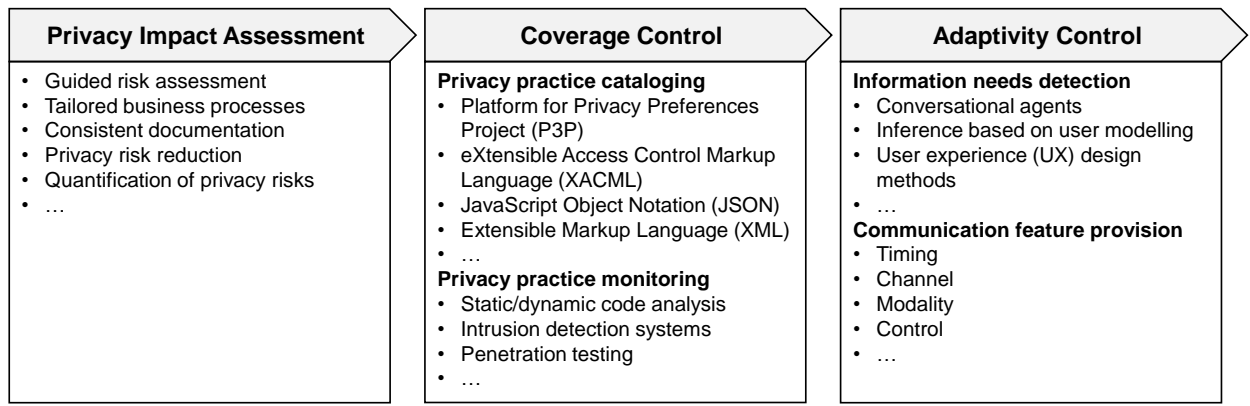

*Figure 3. Overview of three main iterative steps and extant tools and guidance for transparency artifact instantiation.*

**Limitations**

Although TIPP theory is in principle testable, a limitation is that only time tells how well transparency artifacts perform in the real world. To avoid fallacies such as falling for the privacy paradox (Solove 2021), decisive tests of TIPP theory can be done by analyzing transparency artifact instantiations in studies using real tasks, real users, and real systems (Sun and Kantor 2006). This will not only be complex, time-intensive, and costly, but will also raise ethical concerns; after all, the reputation and income of real providers and consumers will be at stake (Culnan 2019). Accordingly, we opted for pure theory development and leave (partial) tests of TIPP theory to future research.

Another limitation of TIPP theory is that we had to focus it on essential aspects for the sake of parsimony; in particular, we restrained ourselves to a higher level of abstraction for the presentation of the inner environment of transparency artifacts and focused the theory development on the consumer-provider dyad. Our goal was to develop an ISDT focused on the design product (Walls et al. 1992, 2004) because TIPP theory can be grounded in a manageable number of kernel theories up to this point. Further refinement of TIPP theory with ISDT focused on context-specific design processes could be performed based on existing research, as outlined in the previous section. In addition, design science research projects targeting uncharted areas where TIPP theory cannot be intuitively applied by practitioners without further refinement via ISDT focused on context-specific design processes (eg, how to effectively map user interaction with an IS to privacy information needs) seem promising for future research.

Another challenge is that theory development is a creative process (Weick 1989). We addressed this challenge by explicating how we instantiated the framework of disciplined imagination, following Weick's suggestion to specify an explicit starting point for the theory development, which "allows other people to begin at the same place and see where their thinking leads them" (1989, p. 529). Any of the thought trial examples in the appendix could be used as starting point.

**Future research**

Opportunities for future research include the (partial) testing of TIPP theory. For instance, it would be interesting to investigate the usefulness of different approaches for implementing coverage and adaptivity feedback loops to find out which approaches work well in different operational settings and for different consumer communities. Thus far, we could not falsify premises or claims of TIPP theory. To further bolster credibility of TIPP theory, we encourage other researchers to challenge and test its premises and claims for further clarification and expansion of the boundaries of TIPP theory. With respect to the consumer perspective, it would be worthwhile to investigate how different types of privacy behaviors result in different information needs. From the provider perspective, it would be interesting to examine which combinations of societal pressures expedite the emergence of useful transparency artifacts. Another interesting question would be whether it is more efficient for providers to operate transparency artifacts on their own or to outsource presentation of the relevant information on privacy practices to specialized third parties operating one-stop shops for TIPP (Schneider and Sunyaev 2016). From a methodological perspective, objective evaluation methods for transparency artifacts should be developed. Reporting standards specifying, for example, the target group of the transparency artifact, features offered by the artifact, the privacy conceptualizations accounted for in the design, and evaluation designs and results, would be helpful for fostering a better understanding of the intended capabilities and limits of transparency artifacts.

**Conclusions**

To contribute to alleviating the privacy challenges resulting from the rising volume of information processing throughout society and to avoid transparency artifact designs where consumers cannot see the 'forest for the trees', useful transparency artifacts must account for the pluralistic nature of privacy and the sociotechnical interplay related to establishing TIPP. This will be helpful for consumers by enabling them to interact with IS in line with their privacy expectations. With respect to TIPP, privacy should not be considered a purely legal, technical, or psychological issue. Privacy is ultimately a social issue, and the design of useful transparency artifacts must account for consumers' evolving and context-dependent privacy expectations and diverse information needs. Most importantly, transparency artifacts must be adaptive to the privacy expectations currently cherished by consumers. Paying too little attention to the sociotechnical perspective may result in technical solutions that outperform privacy notices but still perform poorly with respect to consumers' actual information needs. Many potential avenues for enhancing the design of transparency artifacts remain uncharted. TIPP theory paves the way by offering a theoretical foundation for improved transparency artifact designs from a sociotechnical perspective. We hope that TIPP theory will be useful for IS providers to realize that privacy notices are not a useful transparency artifact and that better artifact designs are necessary to actually establish TIPP. Likewise, public policy initiatives could build on TIPP theory to clarify what they actually call for when demanding transparency and avoid misguided and counterproductive interventions, such as mandatory cookie disclaimers, in the name of transparency. In the end, the nature of privacy is constantly evolving and determined by processes in consumers' minds, which remain unknown. Future transparency artifact designs must become more flexible and adaptive to achieve lasting alignment between technical artifact capabilities and consumers' evolving and context-dependent privacy expectations, so that transparency artifacts can fulfill their fundamental purpose, which is to uphold (and not to subvert) privacy as a social value.

**Acknowledgements**

We would like to thank the three anonymous reviewers, the associated editor, and the senior editor who helped us to strengthen TIPP theory with their helpful and constructive feedback. We would also like to thank the numerous people who supported development of TIPP theory with discussions and friendly reviews of earlier manuscript versions. This work was supported by funding from the topic Engineering Secure Systems of the Helmholtz Association (HGF) and by KASTEL Security Research Labs.

Dinev T, Hart P (2006) An extended privacy calculus model for e-commerce transactions. *Information Systems Research* 17(1):61–80.
Dinev T, McConnell AR, Smith HJ (2015) Informing privacy research through information systems, psychology, and behavioral economics: Thinking outside the "APCO" box. *Information Systems Research* 26(4):639–655.
Donaldson T, Dunfee TW (1994) Toward a unified conception of business ethics: Integrative social contracts theory. *Academy of Management Review* 19(2):252–284.
Donaldson T, Dunfee TW (1995) Integrative social contracts theory: A communitarian conception of economic ethics. *Economics and Philosophy* 11(1):85–112.
Donaldson T, Dunfee TW (1999) *Ties that bind: A social contracts approach to business ethics* (Harvard Business School Press, Boston, MA, USA).
Earp JB, Antón AI, Aiman-Smith L, Stufflebeam WH (2005) Examining internet privacy policies within the context of user privacy values. *IEEE Transactions on Engineering Management* 52(2):227–237.
Elliston FA (1982) Anonymity and whistleblowing. *Journal of Business Ethics* 1(3):167–177.
Feigenbaum J, Ford B (2015) Seeking anonymity in an internet panopticon. *Communications of the ACM* 58(10):58–69.
Feigenbaum J, Freedman MJ, Sander T, Shostack A (2002) Privacy engineering for digital rights management systems. Sander T, ed. *Security and Privacy in Digital Rights Management*. (Springer, Berlin, Heidelberg, Germany), 76–105.
Fuller LL (1969) *The morality of law* 2nd ed. (Yale University Press, New Haven, CT, USA).
Gal-Or E, Gal-Or R, Penmetsa N (2018) The role of user privacy concerns in shaping competition among platforms. *Information Systems Research* 29(3):698–722.
Gerlach JP, Eling N, Wessels N, Buxmann P (2019) Flamingos on a slackline: Companies' challenges of balancing the competing demands of handling customer information and privacy. *Information Systems Journal* 29(2):548–575.
Granados N, Gupta A, Kauffman RJ (2010) Information transparency in business-to-consumer markets: Concepts, framework, and research agenda. *Information Systems Research* 21(2):207–226.
Greenaway KE, Chan YE (2005) Theoretical explanations for firms' information privacy behaviors. *Journal of the Association for Information Systems* 6(6):171–198.
Greenaway KE, Chan YE, Crossler RE (2015) Company information privacy orientation: A conceptual framework. *Information Systems Journal* 25(6):579–606.
Greenleaf G (2014) Sheherezade and the 101 data privacy laws: Origins, significance and global trajectories. *Journal of Law, Information & Science* 23(1):4–49.
Harkous H, Fawaz K, Lebret R, Schaub F, Shin KG, Aberer K (2018) Polisis: Automated analysis and presentation of privacy policies using deep learning. *27th USENIX Security Symposium*. (USENIX Association, Baltimore, MD), 531–548.
Hart S, Ferrara AL, Paci F (2020) Fuzzy-based approach to assess and prioritize privacy risks. *Soft Computing* 24(3):1553–1563.
Hedström P, Swedberg R (1996) Social mechanisms. *Acta Sociologica* 39(3):281–308.
Henriksen-Bulmer J, Faily S, Jeary S (2019) Privacy risk assessment in context: A meta-model based on contextual integrity. *Computers & Security* 82:270–283.
Hoofnagle CJ, Urban JM (2014) Alan Westin's privacy homo economicus. *Wake Forest Law Review* 49:261–317.
Hornby AS (2000) *Oxford Advanced Learner's Dictionary of current English* 6th ed. Wehmeier S, ed. (Oxford University Press, Oxford, UK).
Hornyak R, Rai A, Dong JQ (2020) Incumbent system context and job outcomes of effective enterprise system use. *Journal of the Association for Information Systems* 21(2):364–387.
Hosseini M, Shahri A, Phalp K, Ali R (2018) Engineering transparency requirements: A modelling and analysis framework. *Information Systems* 74(1):3–22.
Hu M (2020) Cambridge Analytica's black box. *Big Data & Society* 7(2):1–6.
Iivari J (2020) A critical look at theories in design science research. *Journal of the Association for Information Systems* 21(3):502–519.

Sheng X, Simpson PM (2014) Effects of perceived privacy protection: Does reading privacy notices matter? *International Journal of Services and Standards* 9(1):19–36.

Simon HA (1996) *The sciences of the artificial* 3rd ed. (MIT Press, Cambridge, MA, USA).

Sinnreich A, Gilbert J (2019) The carrier wave principle. *International Journal of Communication* 13:5816–5840.

Slepchuk AN, Milne GR (2020) Informing the design of better privacy policies. *Current Opinion in Psychology* 31:89–93.

Smith HJ (1993) Privacy policies and practices: Inside the organizational maze. *Communications of the ACM* 36(12):104–122.

Smith HJ, Dinev T, Xu H (2011) Information privacy research: An interdisciplinary review. *MIS Quarterly* 35(4):989–1015.

Smith HJ, Milberg SJ, Burke SJ (1996) Information privacy: Measuring individuals' concerns about organizational practices. *MIS Quarterly* 20(2):167–196.

Soh C, Markus ML, Goh KH (2006) Electronic marketplaces and price transparency: Strategy, information technology, and success. *MIS Quarterly* 30(3):705–723.

Solove DJ (2006) A taxonomy of privacy. *University of Pennsylvania Law Review* 154(3):477–560.

Solove DJ (2021) The myth of the privacy paradox. *George Washington Law Review* 89(1):1–51.

Son JY, Kim SS (2008) Internet users' information privacy-protective responses: A taxonomy and a nomological model. *MIS Quarterly* 32(3):503–529.

Soumelidou A, Tsohou A (2021) Towards the creation of a profile of the information privacy aware user through a systematic literature review of information privacy awareness. *Telematics and Informatics* 61:101592.

Spiekermann S, Korunovska J, Langheinrich M (2019) Inside the organization: Why privacy and security engineering is a challenge for engineers. *Proceedings of the IEEE* 107(3):600–615.

Such JM, Criado N (2018) Multiparty privacy in social media. *Communications of the ACM* 61(8):74–81.

Sun Y, Kantor PB (2006) Cross-evaluation: A new model for information system evaluation. *Journal of the American Society for Information Science and Technology* 57(5):614–628.

Sunyaev A (2020) Critical information infrastructures. *Internet Computing: Principles of distributed systems and emerging internet-based technologies*. (Springer International Publishing, Cham), 339–372.

Sunyaev A, Dehling T, Taylor PL, Mandl KD (2015) Availability and quality of mobile health app privacy policies. *Journal of the American Medical Informatics Association* 22(e1):e28–e33.

Sunyaev A, Kannengießer N, Beck R, Treiblmaier H, Lacity M, Kranz J, Fridgen G, Spankowski U, Luckow A (2021) Token Economy. *Business & Information Systems Engineering* 63(1):457–478.

Sweller J (1988) Cognitive load during problem solving: Effects on learning. *Cognitive Science* 12(2):257–285.

Sweller J, van Merriënboer JJG, Paas F (2019) Cognitive Architecture and Instructional Design: 20 Years Later. *Educational Psychology Review* 31(2):261–292.

Sweller J, van Merriënboer JJG, Paas FGWC (1998) Cognitive architecture and instructional design. *Educational Psychology Review* 10(3):251–296.

Tavani HT (2007) Philosophical theories of privacy: Implications for an adequate online privacy policy. *Metaphilosophy* 38(1):1–22.

Trist E (1981) The evolution of socio-technical systems. *Perspectives in Organization Design and Behavior*. (John Wiley, London, UK), 32–47.

Tsai JY, Egelman S, Cranor L, Acquisti A (2011) The effect of online privacy information on purchasing behavior: An experimental study. *Information Systems Research* 22(2):254–268.

Turner EC, Dasgupta S (2003) Privacy on the Web: An examination of user concerns, technology, and implications for business organizations and individuals. *Information Systems Management* 20(1):8–18.

US Federal Department of Health Education and Welfare (1973) Records, computers and the rights of citizens: Report of the secretary's advisory committee on automated personal data systems. Chapter III. Safeguards for privacy. Retrieved (January 13, 2023), https://epic.org/privacy/hew1973report/c3.htm.